\documentclass[11pt]{article}

\pdfoutput=1

\usepackage{bbm}
\usepackage{multirow}
\usepackage{cite}
\usepackage[dvipsnames]{xcolor}
\usepackage{amssymb,amsmath,amscd}
\usepackage{epsfig}
\usepackage{epstopdf}
\usepackage{latexsym}
\usepackage{graphicx}
\usepackage{booktabs}
\usepackage{bbm}
\usepackage[margin=15pt,small]{caption}
\usepackage{subcaption}
\usepackage{enumitem}
\usepackage{float}
\usepackage[toc]{appendix}
\usepackage[numbers,compress]{natbib}
\usepackage{tikz}
\usepackage[labelfont=bf]{caption}
\usepackage[export]{adjustbox}

\usepackage{braket,cellspace}
\usepackage{mathrsfs}
\usepackage{braket}
\usepackage{chngpage}
\usepackage{tabu}

\allowdisplaybreaks

\newlength{\intwidth}
\DeclareRobustCommand{\fpint}[2]
   {\mathop{%
      \text{%
        \settowidth{\intwidth}{$\int$}%
        \makebox[0pt][l]{\makebox[\intwidth]{$-$}}%
        $\int_{#1}^{#2}$}}}

\newcommand{\mcite}[1]{\mbox{\cite{#1}}}

\usetikzlibrary{positioning}
\definecolor{darkspringgreen}{rgb}{0.09, 0.45, 0.27}

\usepackage{datetime}
\usepackage[
      colorlinks=true,
      linkcolor=darkspringgreen,  
      urlcolor=darkspringgreen,    
      filecolor=darkspringgreen,     
      citecolor=darkspringgreen,
      linktocpage=true,
      pdfstartview=FitV,
      bookmarksopen=true     
      ]{hyperref}

\setlist{nolistsep}
\ifpdf
\fi

\let\oldbibliography\thebibliography
\renewcommand{\thebibliography}[1]{\oldbibliography{#1}
\setlength{\itemsep}{0pt}} 

\numberwithin{equation}{section} 

\begin{document}  

\begin{titlepage}

\begin{center} 

\vspace*{20mm}

{\LARGE \bf 
$T\bar{T}$ deformation of random matrices
}

\bigskip
\bigskip
\bigskip
\bigskip

{\bf Felipe Rosso \\ }
\vskip 2mm
Department of Physics and Astronomy 
\\
University of Southern California \\
Los Angeles, CA 90089, USA  \\
\vskip 2mm
Kavli Institute for Theoretical Physics \\
University of California\\
 Santa Barbara, CA 93106, USA \\
\bigskip
\tt{
felipero@usc.edu}  \\

\end{center}

\bigskip

\begin{abstract}
\noindent We define and study the $T\bar{T}$ deformation of a random matrix model, showing a consistent definition requires the inclusion of both the perturbative and non-perturbative solutions to the flow equation. The deformed model is well defined for arbitrary values of the coupling, exhibiting a phase transition for the critical value in which the spectrum complexifies. The transition is between a single and a double-cut phase, typically third order and in the same universality class as the Gross-Witten transition in lattice gauge theory. The $T\bar{T}$ deformation of a double scaled model is more subtle and complicated, and we are not able to give a compelling definition, although we discuss obstacles and possible alternatives. Quantitative comparisons with finite cut-off Jackiw-Teitelboim gravity are presented.
\end{abstract}

\vfill

\end{titlepage}


\newpage

\setcounter{tocdepth}{2}
\tableofcontents

\section{Introduction}
\label{sec:1}

The $T\bar{T}$ deformation was first introduced as a deformation of two-dimensional quantum field theories (QFTs) by an irrelevant operator built from the stress tensor $T_{\mu \nu}$ \cite{Smirnov:2016lqw,Cavaglia:2016oda} (see \cite{Jiang:2019hxb} for a review). It distinguishes itself from other irrelevant deformations given that several observables of the deformed theory (energy spectrum, partition function and S-matrix, among others) can be computed exactly and unambiguously~\mcite{Smirnov:2016lqw,Cavaglia:2016oda,Jiang:2019hxb,Zamolodchikov:2004ce,Dubovsky:2012wk,Dubovsky:2017cnj,Datta:2018thy,Aharony:2018bad}. This is quite surprising, as general arguments in renormalization theory imply observables in the deformed theory are not well defined, as they require an infinite number of counter terms. Encouraged by these interesting results in two-dimensional QFTs, the $T\bar{T}$ deformation has been generalized to other setups, including higher dimensional QFTs~\cite{Bonelli:2018kik,Taylor:2018xcy,Hartman:2018tkw}, quantum mechanics~\cite{Gross:2019ach,Gross:2019uxi}, spin chains~\cite{Bargheer:2008jt,Bargheer:2009xy,Marchetto:2019yyt,Pozsgay:2019ekd} and holography~\cite{McGough:2016lol,Guica:2019nzm,Kraus:2018xrn,Iliesiu:2020zld}. This provides new perspectives from which to study certain aspects of the~${T\bar{T}}$ deformation that are not simple to understand in its original formulation for two-dimensional QFTs. 

As an example, let us consider the $T\bar{T}$ deformation of a quantum mechanical system characterized by a Hamiltonian operator $H$. The deformation defined in \cite{Gross:2019ach} and reviewed in section \ref{sec:2}, is implemented by deforming the Hamiltonian operator $H$ according to the following flow equation
\begin{equation}\label{eq:30}
\partial_\lambda H(\lambda)=\frac{2H(\lambda)^2}{1-4\lambda H(\lambda)}
\qquad \quad \Longrightarrow \qquad \quad
H_\pm(\lambda)=
\frac{1\pm\sqrt{1-8\lambda H}}{4\lambda}\ ,
\end{equation}
where $\lambda \in \mathbb{R}$ is the deformation parameter. While there are two branches $H_\pm(\lambda)$ that solve the flow equation, only the negative branch is perturbatively connected to the undeformed theory, i.e.~${H_-(\lambda=0)=H}$, meaning $H_+$ is the non-perturbative branch.\footnote{In both branches, the integration constant is fixed to the same value so that the two solutions are smoothly connected.} We immediately identify an issue with this deformation, since nothing prevents the argument in the square root going negative, resulting in a complex energy spectrum of the deformed theory. This issue also arises in the $T\bar{T}$ deformation of two-dimensional QFTs and it is currently unclear how one should deal with it. Should we restore unitarity by introducing a truncation of the spectrum of $H$ that ensures $H_\pm^\dagger=H_\pm$? If so, is this procedure unique? Or maybe we should accept a non-unitary deformed theory?

One of the goals of this work is to address these questions by defining and studying the $T\bar{T}$ deformation in (perhaps) the simplest setup: a random matrix model. We shall mainly focus on an ensemble of Hermitian square matrices $M$ of dimension $N$, weighted by a probability measure determined by a potential $V(M)$ according to $dMe^{-\frac{N}{\gamma}\,{\rm Tr}\,V(M)}$. The expectation value of any matrix observable $\mathcal{O}$ is defined as
\begin{equation}\label{eq:97}
\langle
\mathcal{O}
\rangle\equiv
\frac{1}{\mathcal{Z}}
\int dM
\mathcal{O}\,
e^{-\frac{N}{\gamma}\,{\rm Tr}\,V(M)}\ ,
\qquad \qquad
\mathcal{Z}\equiv
\int dMe^{-\frac{N}{\gamma}\,{\rm Tr}\,V(M)}\ ,
\end{equation}
where $\gamma\in \mathbb{R}$. One of the central observables is the spectral density
\begin{equation}
\rho(E)=
\frac{1}{N}
\Big\langle
\sum_{i=1}^N
\delta(E-\alpha_i)
\Big\rangle\ ,
\end{equation}
which characterizes the average distribution of eigenvalues $\alpha_i\in \mathbb{R}$ of the matrix $M$. 

Starting from the flow equation in (\ref{eq:30}) we show in section \ref{sec:3} that in order to have a consistent definition of the $T\bar{T}$ deformation of a random matrix model we must \textit{necessarily} include the contribution from both the perturbative and non-perturbative solutions $H_\pm$. We then show the deformation is very naturally defined in terms of the potential $V(M)$ according to
\begin{equation}\label{eq:31}
T\bar{T}{\rm \,\,deformation:}
\qquad \quad
V_\lambda(M)=c_\lambda V(M-2\lambda M^2)
\ ,
\end{equation}
where $c_\lambda=1/2$ except for $c_{\lambda=0}=1$. This provides the perfect setup for studying the behavior of the system when~$\lambda>\lambda_c$. As we take~$\lambda$ across its critical value~$\lambda_c$ we observe a shift in the extrema of the potential~$V_\lambda(M)$. This triggers a phase transition in the large $N$ spectral density $\rho_0(E)$, which goes from being supported in two disjoint intervals (double-cut) to a single one (single-cut), see figures~\ref{fig:1} and~\ref{fig:8}. This type of transition is very familiar to random matrix models. Starting from a typical potential $V(M)$ the phase transition at $\lambda_c$ is of third order and in the same universality class as the Gross-Witten transition in two-dimensional gauge theory \cite{Gross:1980he}. Crucially, there is no need to introduce a truncation of the spectrum or any other ad-hoc procedure, as the matrix model has the appropriate structure to deal with the phase transition in a unique and natural way. The reminder of section \ref{sec:3} is devoted to the study of additional features of the deformation (\ref{eq:31}). This includes some remarkable stability considerations, the deformation of critical potentials and the definition of the deformation for unitary matrix models (see equation (\ref{eq:52})). 


In subsection \ref{subsec:5.1} we study the $T\bar{T}$ deformation of double scaled random matrix models. A double scaled model is obtained from a \textit{critical} potential $V(M)$ by simultaneously taking the limit~${N\rightarrow \infty}$ and~${\gamma\rightarrow 1}$ in a particular way (see \cite{Ginsparg:1993is,DiFrancesco:1993cyw} for reviews). This procedure has the effect of ``zooming in" to the edge of the eigenvalue spectrum~$\rho(E)$, capturing universal physics while disregarding non-universal features of~$\rho(E)$ away from the edge.\footnote{Double scaled models can also be defined independently of any matrix model, in terms of a topological expansion associated to an algebraic curve \cite{Eynard:2007kz}.} Defining the deformation of a double scaled model turns out being much more complicated than that of an ordinary matrix model with no double scaling. The fundamental obstacle is that applying the double scaling limit on a matrix model is not an invertible procedure, i.e. given a double scaled model there is not a unique potential $V(M)$ associated to it. This hinders the utility of the simple definition of the deformation given in (\ref{eq:31}). 
That being said, we are still able to give some partial definitions and study certain aspects of the $T\bar{T}$ deformation of double scaled models, that we hope can set the stage for future investigations.

Our interest in double scaled models arises from interesting connections between the $T\bar{T}$ deformation and finite cut-off AdS holography \cite{McGough:2016lol}. The thermal partition function of the $T\bar{T}$ deformation~(\ref{eq:30}) of the Schwarzian quantum mechanics has been recently reproduced from the finite cut-off Jackiw-Teitelboim~(JT) gravity disc partition function \cite{Iliesiu:2020zld} (see also \cite{Stanford:2020qhm}). Since higher topology contributions in ordinary JT gravity are captured by a double scaled model \cite{Saad:2019lba}, is there a matrix model that captures higher topology contributions in finite cut-off JT gravity? To answer this question we can compute higher genus finite cut-off partition functions using the decomposition of surfaces with constant negative surfaces developed in \cite{Saad:2019lba}, in terms of the ``trumpet" geometry. Using the finite cut-off trumpet partition function of~\cite{Iliesiu:2020zld}, in subsection \ref{subsec:5.2} we calculate the leading genus two boundary partition function and show it is not compatible with matrix model predictions.\footnote{Since the leading expectation value of double trace operators in a matrix model are universal, i.e. independent of the particular details of the model, the comparison can be made in full generality.} This shows the (by now) standard approach for computing higher genus Euclidean partition function of~\cite{Saad:2019lba}, might not be useful when applied to finite cut-off JT gravity, and instead a different procedure has to be developed.


\vspace{7pt}

\noindent \textbf{Note:} While this work was near completion, reference \cite{Gorsky:2020qge} appeared. That work studies the $T\bar{T}$ deformation of two-dimensional large $N$ Yang-Mills theory and also finds a phase transition by accounting for the non-perturbative branch solving the flow equation. Given the relation between large $N$ gauge theory and matrix models \cite{Gross:1980he}, perhaps a connection can be made between the results in this work and \cite{Gorsky:2020qge} (see also \cite{Santilli:2020qvd,Santilli:2018xux}).

\section{Quantum mechanics}
\label{sec:2}

In this section we start by reviewing the $T\bar{T}$ deformation proposed in \cite{Gross:2019ach} (see also \cite{Gross:2019uxi,Chakraborty:2020xwo}) for a quantum mechanical system. This sets the stage for the definition of the deformation of a random matrix model in the next section. Consider a quantum mechanical system characterized by a hermitian Hamiltonian operator $H$ that satisfies
\begin{equation}\label{eq:63}
H\ket{\psi_E}=E\ket{\psi_E}\ ,
\qquad \qquad
E\in \mathcal{S}(H)\ ,
\end{equation}
where $\ket{\psi_E}$ and $\mathcal{S}(H)$ are the eigenstates and spectrum of $H$ respectively. Let us assume the spectrum of this system is supported on the finite interval $\mathcal{S}(H)\in[0,E_0]$. Our focus is on the thermal partition function, defined as
\begin{equation}\label{eq:17}
Z(\beta)\equiv {\rm Tr}
\big(e^{-\beta H}\big)=
\int_0^{E_0}
dE\,
\rho(E)e^{-\beta E}\ ,
\end{equation}
where the spectral density $\rho(E)$ determines the degeneracy of the eigenstates~$\ket{\psi_E}$.

The $T\bar{T}$ deformation proposed in \cite{Gross:2019ach} (obtained from dimensional reduction of the original two-dimensional deformation \cite{Smirnov:2016lqw,Cavaglia:2016oda}) is defined from the flow equation for the deformed Hamiltonian given in (\ref{eq:30}). There are two distinct branches that solve the differential equation, so that the deformed eigenvalues are given by
\begin{equation}\label{eq:71}
E_\pm(\lambda,E)=\frac{1\pm \sqrt{1-8\lambda E}}{4\lambda}\in \mathbb{R}
\qquad \Longleftrightarrow \qquad
\lambda\le \lambda_c(E_0)\equiv \frac{1}{8E_0}\ .
\end{equation}
In this section we shal restrict to $\lambda\le \lambda_c$ so that $E_\pm(\lambda,E)\in \mathbb{R}$. While the perturbative branch~${E_-(\lambda,E)}$ satisfies $E_-(0,E)=E$, the non-perturbative branch $E_+(\lambda,E)$ diverges as $\lambda\rightarrow 0$. Despite this singular behavior, we shall see that both branches solving the flow equation play a crucial role when defining the deformation for a random matrix model. Note that the eigenstates~$\ket{\psi_E}$ are not modified by the deformation.

We are interested in studying the effect of the deformation on the spectral density $\rho(E)$ appearing in the thermal partition function (\ref{eq:17}). In doing so, we make the distinction between two different approaches, that involve considering a single or both branches in (\ref{eq:71}).

\paragraph{Single branch:}

The most naive approach is to ignore the non-perturbative branch $H_+$ and only include the perturbative solution~$H_-$. The deformed partition function in this case is given by
\begin{equation}\label{eq:20}
Z_\lambda(\beta)
\equiv
{\rm Tr}\big(e^{-\beta H_-}\big)=
\int_0^{E_0}dE\,
\rho(E)e^{-\beta E_-(\lambda,E)}\ .
\end{equation}
The density $\rho(E)$ appearing in (\ref{eq:20}) is the same as in the undeformed theory since for any given value of $E$, the eigenstates $\ket{\psi_E}$ are unchanged. The deformed spectral density $\rho_\lambda(E)$ is obtained by changing the variables of the integral in (\ref{eq:20}) so that we get the standard Boltzmann factor $e^{-\beta E}$. Doing so, we find
\begin{equation}\label{eq:21}
\rho_\lambda(E)=
(1-4\lambda E)\rho(E-2\lambda E^2)
\times \textbf{1}_{[0,E_-(\lambda,E_0)]}\ ,
\qquad
\lambda\le \lambda_c\ ,
\end{equation} 
where the indicator function is given by
\begin{equation}\label{eq:23}
\textbf{1}_{A}=
\begin{cases}
\begin{aligned}
\,\,\,&1\,\,\, \ ,\,\,\, E\in A\ , \\
\,\,\,&0\,\,\, \ , \,\,\, E\notin A\ ,
\end{aligned}
\end{cases}
\end{equation}
for any set $A$. The right edge of the spectrum is determined by the negative branch $E_-$ in (\ref{eq:71}) evaluated at $E_0$. Note the prefactor $(1-4\lambda E)$ in (\ref{eq:21}) is non-negative in the support of $\rho_\lambda(E)$.

\paragraph{Both branches:}

A second approach involves including both branches that solve the flow equation~(\ref{eq:71}), so that the thermal partition function is given by 
\begin{equation}\label{eq:25}
\widetilde{Z}_\lambda(\beta)\equiv
c_\lambda
{\rm Tr}\big(
e^{-\beta H_-}+
e^{-\beta H_+}
\big)=
c_\lambda
\int_0^{E_0}
dE\,\rho(E)
\big(
e^{-\beta E_-(\lambda,E)}+
e^{-\beta E_+(\lambda,E)}
\big)	\ ,
\end{equation}
where we add a tilde to differentiate from the previous prescription. The normalization constant $c_\lambda$ is defined as
\begin{equation}
c_\lambda=\begin{cases}
\,\,\,\,\,\,1\quad \,\,\, \ , \quad \lambda=0\ , \\
\,\,\,1/2 \quad \ , \quad \lambda \neq  0\ ,
\end{cases}
\end{equation}
which ensures a proper normalization for all $\lambda$. The idea of including both branches in this way was first explored in \cite{Iliesiu:2020zld} when studying JT gravity with a finite cut-off, with the important difference that a different (and arbitrary) spectral density~$\rho_+(E)$ was considered for the non-perturbative branch~$E_+$. From our perspective, we have the undeformed spectral density~$\rho(E)$ for both branches since both traces in~(\ref{eq:25}) are computed with respect to the undeformed eigenstates $\ket{\psi_E}$. Changing coordinates in each term so that we get the standard Boltzmann factor, we identify the deformed spectral density as
\begin{equation}\label{eq:22}
\widetilde{\rho}_\lambda(E)=
c_\lambda
|1-4\lambda E|
\rho(E-2\lambda E^2)
\times 
\big[
\textbf{1}_{[0,E_-(\lambda,E_0)]}+
\textbf{1}_{[E_+(\lambda, E_0),1/2\lambda]}
\big]\ ,
\qquad
\lambda\le \lambda_c\ .
\end{equation}
The absolute value in the prefactor $|1-4\lambda E|$ arises from the different change of variables involved in each term in (\ref{eq:25}). 

\begin{figure}
\centering
\includegraphics[scale=0.45]{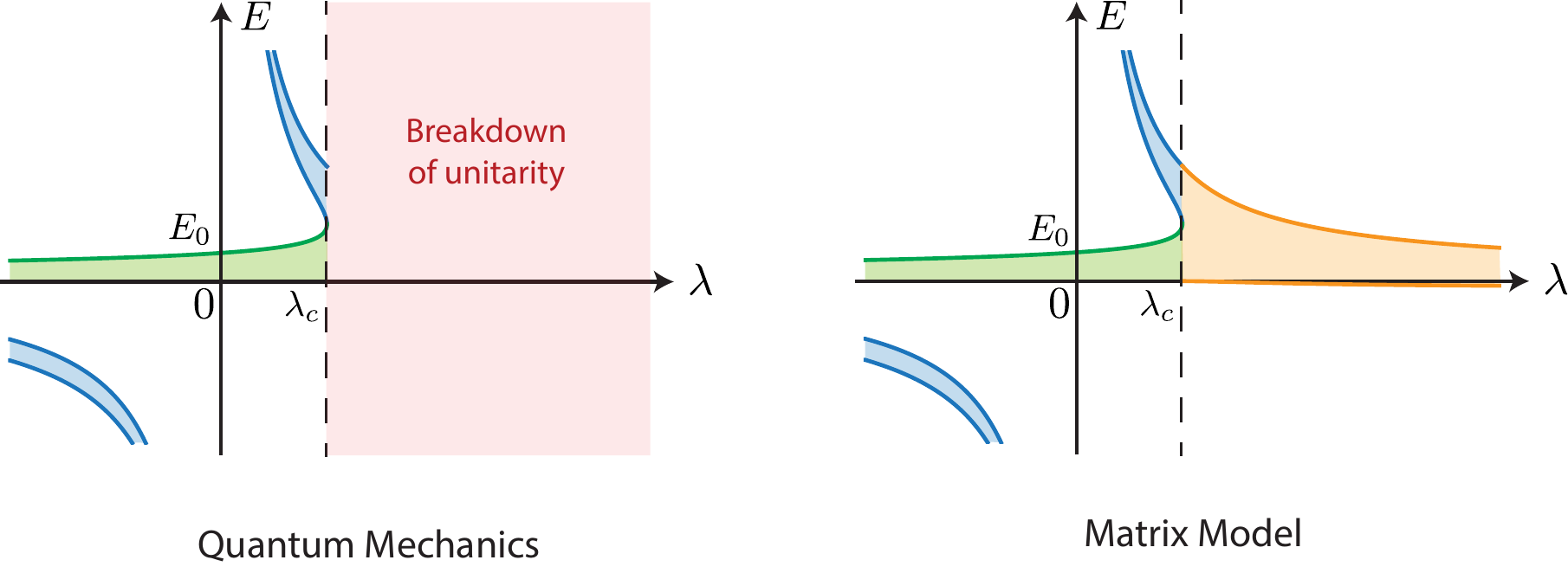}
\caption{On the left we plot the support of $\widetilde{\rho}_\lambda(E)$ in (\ref{eq:22}) for the quantum mechanics $T\bar{T}$ deformation. Green and blue shaded regions correspond to contributions coming from the perturbative and non-perturbative branches respectively. Since the energy spectrum complexifies, we are unable to go beyond~$\lambda_c$. On the right, we show the corresponding diagram obtained by $T\bar{T}$ deformation the matrix model studied in subsection \ref{sub:3.1}. In this case, we can go through and beyond the phase transition.}
\label{fig:6}
\end{figure}

Comparing with $\rho_\lambda(E)$ in (\ref{eq:21}) we see the spectral density gets an additional contribution from the non-perturbative branch. The support of the spectral density is much more interesting in this case, as can be seen from the left diagram in figure \ref{fig:6}. The green and blue regions indicate the support of~$\widetilde{\rho}_\lambda(E)$ arising from contributions of the perturbative and non-perturbative branches respectively. As $\lambda\rightarrow 0$ the contribution of the positive branch $E_+$ (blue in figure \ref{fig:6}) goes to zero, while the negative branch $E_-$ (green in figure \ref{fig:6}) goes to the undeformed density $E\in[0,E_0]$. As we approach $\lambda_c$ the square root in both branches (\ref{eq:71}) vanishes and the spectral density becomes supported on a single interval
\begin{equation}
\widetilde{\rho}_{\lambda_c}(E)=
c_{\lambda_c}
|1-4\lambda_c E|
\rho(E-2\lambda_c E^2)
\times 
\textbf{1}_{[0,4E_0]}\ .
\end{equation}
This hints towards a phase transition, in which the spectral density goes from a double to a single-cut phase as $\lambda\ge \lambda_c$. While from the perspective of the quantum mechanics there is no clear way of going beyond this transition, we shall show how the matrix model is naturally equipped to deal with it. We will be able to continue $\widetilde{\rho}_\lambda(E)$ beyond $\lambda_c$ and find its support is given by the right diagram in figure \ref{fig:6}.

\section{Random matrix models}
\label{sec:3}

In this section we define and study the $T\bar{T}$ deformation of a random matrix model. Let us start with a short introduction to random matrix models, for reviews see \cite{DiFrancesco:1993cyw,Ginsparg:1993is,Eynard:2015aea}. Consider an ensemble of Hermitian matrices $M$ of dimension $N$, weighted by a probability measure determined by a potential~$V(M)$ according to $dMe^{-\frac{N}{\gamma}\,{\rm Tr}\,V(M)}$, where $dM$ is the $U(N)$ invariant measure. The expectation value of matrix model observables $\mathcal{O}$ are computed according to (\ref{eq:97}). Two useful observables are the spectral density $\rho(E)$ and resolvent $R(z)$, defined as
\begin{equation}\label{eq:12}
\rho(E)\equiv \frac{1}{N}
\Big\langle
\sum_{i=1}^N\delta(E-\alpha_i)
\Big\rangle\ ,
\qquad \qquad
R(z)\equiv \frac{1}{N}
\Big\langle
\sum_{i=1}^N
\frac{1}{z-\alpha_i}
\Big\rangle=\frac{1}{z}+\mathcal{O}(1/z^2)\ ,
\end{equation}
where $\alpha_i$ are the eigenvalues of the matrix $M$ and $R(z)$ is an analytic function in $z\in \mathbb{C}\setminus \lbrace{\rm supp\,\,\rho}\rbrace$. These two quantities are related via the following transformations
\begin{equation}\label{eq:46}
\rho(E)=\lim_{\epsilon\rightarrow 0}
\frac{R(E-i\epsilon)-R(E+i\epsilon)}{2\pi i}\ ,
\qquad \qquad
R(z)=\int_{\rm supp\,\,\rho}
\frac{\rho(E)}{z-E}dE\ .
\end{equation}

We study observables in the large $N$ limit, where we add the subscript zero to differentiate from the finite $N$ quantities, e.g. $R_0(z)\equiv \lim_{N\rightarrow \infty}R(z)$. In this limit the resolvent satisfies a simple algebraic equation that can be solved and written as (e.g. see section 3 of \cite{Eynard:2015aea})
\begin{equation}\label{eq:62}
R_0(z)=\frac{1}{2}
\left[
V'(z)-\sqrt{V'(z)^2-4P_0(z)}
\right]\ .
\end{equation}
Restricting ourselves to polynomial potentials, the function $P_0(z)$ is also a polynomial. Since the argument in the square root in (\ref{eq:62}) is also a polynomial, we can factorize it in terms of its even and odd zeros as $V'(z)^2 -4P_0(z)\equiv h(z)^2\sigma(z)$. We can then use (\ref{eq:46}) to write the large $N$ spectral density as
\begin{equation}\label{eq:61}
\rho_0(E)=
\frac{1}{2\pi}
|h(E)|\sqrt{-\sigma(E)}
\times \textbf{1}_{\sigma(E)< 0}\ .
\end{equation}
The support of the equilibrium spectral density is determined by the function $\sigma(E)$, that can be written as $\sigma(E)=\prod_{i=1}^{2s}(E-a_i)$. The parameters $a_i\in \mathbb{R}$ determine the edges of the spectrum, which must be real since the model is built from Hermitian matrices. The coefficients $a_i$ together with the polynomial $h(E)$ are determined by the potential $V(M)$. For the single cut case the spectral density is supported in a single interval, so that $\sigma(E)=(E-a_-)(E-a_+)$. We can determine $h(E)$ by requiring the resolvent has the appropriate large $z$ limit (\ref{eq:12}). This gives the following condition
\begin{equation}\label{eq:16}
h(z)={\rm Pol}\left[
\frac{V'(z)}{\sqrt{(z-a_-)(z-a_+)}}
\right]\ ,
\end{equation}
where ${\rm Pol}[\,\cdot\,]$ is the polynomial contribution in  $z$ obtained from expanding around $z\rightarrow +\infty$. The values of $a_\pm$ are obtained from
\begin{equation}\label{eq:68}
V'(z)={\rm Pol}\left[
h(z)\sqrt{\sigma(z)}
\right]\ ,
\qquad \qquad
\underset{z=+\infty}{\rm Res}
\left[
h(z)\sqrt{\sigma(z)}
\right]=-2\ ,
\end{equation}
where the second condition is equivalent to requiring the spectral density is properly normalized. 

\subsection{Definition and phase transition}
\label{sub:3.1}

Let us now show how we can implement the $T\bar{T}$ deformation of a random matrix model. Our starting point is a potential $V(x)$ associated to a single-cut spectral density, that without loss of generality we can take as
\begin{equation}\label{eq:58}
\rho_0(E)=\frac{1}{2\pi}|h(E)|\sqrt{E(E_0-E)}
\times \textbf{1}_{[0,E_0]}\ .
\end{equation}
The polynomial $h(E)$ is related to the potential $V(x)$ through the first identity in (\ref{eq:68}). To write this explicitly, we use the following large $z$ expansion
\begin{equation}
h(z)\sqrt{z(z-E_0)}=
\sum_{n=0}^ph_n
\sum_{m=0}^{\infty}
\binom{1/2}{m}(-E_0)^m z^{n+1-m}\ ,
\end{equation}
where $h_n$ are the coefficients of the polynomial $h(E)$ of order $p$. Using this in (\ref{eq:68}) we get
\begin{equation}\label{eq:8}
V'(z)=
\sum_{n=0}^ph_n
\sum_{m=0}^{n+1}
\binom{1/2}{m}(-E_0)^m z^{n+1-m}\ .
\end{equation}
This relation determines the potential necessary to generate any single-cut large $N$ spectral density given by~(\ref{eq:58}). A potential $V(z)$ is said to be critical (see section 6.5 in \cite{Akemann:2011csh}) if the associated polynomial~$h(E)$ vanishes in the support of the spectral density~$\rho_0(E)$, which is called singular. For~(\ref{eq:58}) this corresponds to~$h(E)$ having a zero in the region~$E\in[0,E_0]$. 

\paragraph{Singe branch:}

Let us start by considering the case in which we only include the perturbative branch $E_-(\lambda,E)$, so that the deformed spectral density for $\lambda\le \lambda_c$ is given by (\ref{eq:21}). Applying this to the matrix model large $N$ density (\ref{eq:58}) we find
\begin{equation}\label{eq:57}
\rho_\lambda(E)=
\frac{1}{2\pi}
|h_\lambda(E)|
\sqrt{-\sigma_\lambda(E)}
\times \textbf{1}_{[0,E_-(\lambda,E_0)]}\ ,
\qquad  \quad\lambda\le \lambda_c\ ,
\end{equation}
where we have identified
\begin{equation}\label{eq:64}
\begin{cases}
\begin{aligned}
\,\,\sigma_\lambda(E)&=4\lambda^2
E(E-E_-(\lambda,E_0))(E-E_+(\lambda,E_0))\left(E-1/2\lambda\right)\ , \\[3pt]
h_\lambda(E)&=
(1-4\lambda E)h(E-2\lambda E^2)\ .
\end{aligned}
\end{cases}
\end{equation}
Note that we can include $(1-4\lambda E)$ inside the absolute value since it is positive in the range~${E\in[0,E_-(\lambda,E_0)]}$. To identify (\ref{eq:57}) as the leading order spectral density of a matrix model it must take the general form given in (\ref{eq:61}). While the functional form is appropriately given by the functions $h_\lambda(E)$ and $\sigma_\lambda(E)$, there is an issue with the support of the spectral density, given that the polynomial $\sigma_\lambda(E)$ contains four roots instead of two. More precisely, the issue arises due to the following discrepancy in the indicator functions
\begin{equation}\label{eq:1}
\textbf{1}_{\sigma_\lambda(E)<0}=
\textbf{1}_{[0,E_-(\lambda,E_0)]}+
\textbf{1}_{[E_+(\lambda,E_0),1/2\lambda]}
\neq 
\textbf{1}_{[0,E_-(\lambda,E_0)]}\ .
\end{equation}
As a result, the deformed spectral density that only incorporates the perturbative branch $E_-(\lambda,E)$ cannot be written in the form (\ref{eq:61}) and therefore interpreted as coming from a random matrix model,~i.e.
\begin{equation}\label{eq:98}
\rho_\lambda(E)\neq 
\frac{1}{2\pi}
|h_\lambda(E)|
\sqrt{-\sigma_\lambda(E)}
\times \textbf{1}_{\sigma_\lambda(E)<0}\ .
\end{equation}

\paragraph{Both branches:}

This suggests we study the deformation of the spectral density given in (\ref{eq:22}), that includes both branches $E_\pm(\lambda,E)$ solving the flow equation (\ref{eq:71}). Applying the deformation to the single cut density in (\ref{eq:58}) we find
\begin{equation}\label{eq:66}
\widetilde{\rho}_\lambda(E)=
\frac{1}{2\pi}
\big|\widetilde{h}_\lambda(E)\big|
\sqrt{-\widetilde{\sigma}_\lambda(E)}
\times
\underbrace{\big[
\textbf{1}_{[0,E_-(\lambda,E_0)]}+
\textbf{1}_{[E_+(\lambda, E_0),1/2\lambda]}
\big]}_{\textbf{1}_{\widetilde{\sigma}_\lambda(E)<0}}\ ,
\qquad \quad \lambda\le \lambda_c\ .
\end{equation}
The crucial difference is in the indicator functions, that in this case appropriately combine to yield~${\textbf{1}_{\widetilde{\sigma}_\lambda(E)<0}}$, where we have defined
\begin{equation}
\begin{cases}
\begin{aligned}
\,\,\widetilde{\sigma}_\lambda(E)&=4\lambda^2
E(E-E_-(\lambda,E_0))(E-E_+(\lambda,E_0))\left(E-1/2\lambda\right)\ , \\[3pt]
\widetilde{h}_\lambda(E)&=c_\lambda
(1-4\lambda E)h(E-2\lambda E^2)\ .
\end{aligned}
\end{cases}
\end{equation}
Note the difference in the normalization constant $c_\lambda$ with respect to (\ref{eq:64}). This shows that in order to define the $T\bar{T}$ deformation of a random matrix model we must \textit{necessarily} include the contributions from both branches, as including a single one is inconsistent (\ref{eq:98}). 

To provide a standard definition of the deformation to all order in $1/N$, we need to derive a formula for the potential~${V_\lambda(x)}$. This can be obtained from the spectral density (\ref{eq:66}) and the first relation in (\ref{eq:68}). To do so, we use the following large $z$ expansion
\begin{equation}
\widetilde{h}_\lambda(z)\sqrt{\widetilde{\sigma}_\lambda(z)}=
c_\lambda
\sum_{n=0}^p
h_n 
\sum_{m=0}^{\infty}
\binom{1/2}{m}
(-E_0)^m
(1-4\lambda z)
(z-2\lambda z^2)^{n+1-m}\ .
\end{equation}
Using this in (\ref{eq:68}) we find the following expansion for the deformed potential 
\begin{equation}\label{eq:19}
V'_\lambda(z)=
c_\lambda
(1-4\lambda z)
\sum_{n=0}^p
h_n 
\sum_{m=0}^{n+1}
\binom{1/2}{m}
(-E_0)^m
(z-2\lambda z^2)^{n+1-m}\ .
\end{equation}
While this expression is quite complicated, it greatly simplifies after using the relation satisfied by the undeformed potential in (\ref{eq:8}). This results in the following simple formula for the deformed potential
\begin{equation}\label{eq:81}
V_{\lambda}(x)=
c_\lambda
V(x-2\lambda x^2)\ ,
\end{equation}
where the prefactor $(1-4\lambda z)$ in (\ref{eq:19}) arises from the chain rule after taking the derivative. This provides a simple and natural way of defining the $T\bar{T}$ deformation of a Hermitian random matrix model.

Let us analyze some general features of the deformation formula for the potential. If the undeformed potential is stable, i.e. if ${V(x\rightarrow \pm \infty)=+\infty}$, the deformed potential is also stable for \textit{arbitrary} real values of $\lambda$. This is quite different from the definition of the deformation in quantum mechanics, which only makes sense for $\lambda\le \lambda_c$. Even more, if we start from an unstable potential whose leading behavior is given by $V(x)=-x^{2n+1}+\cdots$, the $T\bar{T}$ deformation cures the instability since the large $x$ behavior changes to $V_{\lambda}(x)= (2\lambda x)^{4n+2}+\cdots$. We shall later consider some examples where we observe this feature explicitly.

Let us now assume the undeformed potential is stable and has a single extremum at $x=x_c$ that is also a minimum. Since the equilibrium spectral density is supported on $E\in[0,E_0]$, the minimum must be located in the same interval $x_c\in[0,E_0]$. The critical points of the deformed potential can be readily computed as
\begin{equation}\label{eq:82}
V'_{\lambda}(x)=0
\qquad \Longleftrightarrow \qquad
x_\pm=\frac{1\pm\sqrt{1-8\lambda x_c}}{4\lambda}\ ,
\qquad
x_1=\frac{1}{4\lambda}\ .
\end{equation}
Using $V_\lambda(\pm \infty)=+\infty$ together with the fact that $x_1$ is always in between the other critical points~$x_\pm$, we conclude that when $x_\pm\in \mathbb{R}$ the points $x_\pm$ correspond to local minima and $x_1$ to a maximum. However, when $x_\pm$ become complex, $x_1$ is the only real critical point which must therefore be a minimum. The shift between these two regimes induces a phase transition in the spectral density, that for finite positive $\lambda$ goes from a double to a single-cut phase. Due to the breakdown of the formula for $\widetilde{\rho}_\lambda(E)$ in (\ref{eq:66}) for~${\lambda>\lambda_c}$, we identify the location of the transition at~${\lambda_c=1/8E_0}$, so that the full spectral density is given by
\begin{equation}\label{eq:83}
\widetilde{\rho}_\lambda(E)=
\frac{1}{2\pi}\times 
\begin{cases}
\begin{aligned}
\,\,\,\big|\widetilde{h}_\lambda(E)\big|
\sqrt{-\widetilde{\sigma}_\lambda(E)}
\times
\big[
\textbf{1}_{[0,E_-(\lambda,E_0)]}+
\textbf{1}_{[E_+(\lambda,E_0),1/2\lambda]}
\big]&\ ,
\qquad
\lambda\le \lambda_c\ ,\\[8pt]
|\bar{h}_\lambda(E)|\sqrt{(E-a_-)(a_+-E)}
\times
\textbf{1}_{[a_-,a_+]}&\ ,
\qquad
\lambda\ge \lambda_c\ .
\end{aligned}
\end{cases}
\end{equation} 
The single-cut spectral density after the transition is characterized by the polynomial $\bar{h}(E)$ and the end points $a_\pm$. As we shall shortly show in an example, these are easily computed from the deformed potential $V_\lambda(x)$ using the conditions in (\ref{eq:16}) and (\ref{eq:68}). Overall, the matrix model is naturally equipped to deal with the phase transition in a unique way, from a double to a single-cut phase. Other methods one could consider, like truncating the spectrum in order to restore unitarity, are easily shown to be inconsistent from the matrix model perspective. 

\begin{figure}
\centering
\qquad \qquad \quad
\includegraphics[scale=0.38]{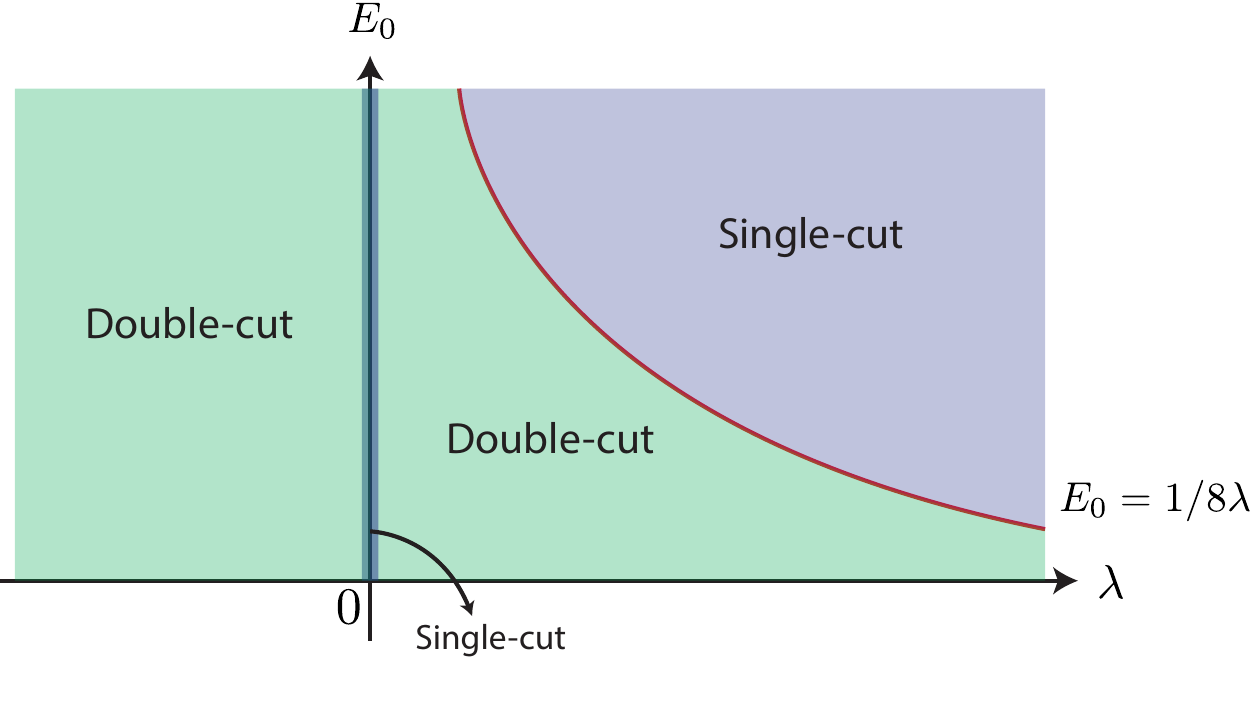}
\caption{Phase diagram in parameter space $(\lambda,E_0)$ showing the transition at $\lambda_c=1/8E_0$ between the single and double-cut spectral densities $\widetilde{\rho}_\lambda(E)$ in (\ref{eq:83}).}\label{fig:1}
\end{figure}

The phase diagram in the parameter space $(\lambda,E_0)$ is sketched in figure \ref{fig:1}, where the two phases are divided by the curve $\lambda_c=1/8E_0$. It is interesting to study the spectral density at criticality, that is given by
\begin{equation}\label{eq:29}
\widetilde{\rho}_{\lambda_c}(E)=
\frac{c_\lambda}{4\pi}
\big|
h(E-2\lambda_c E^2)
\big|
\left(\frac{E-2E_0}{2E_0}\right)^2
\sqrt{
E(4E_0-E)}
\times
\textbf{1}_{[0,4E_0]}
\ .
\end{equation}
Due to the factor $(E-2E_0)^2$ this spectral density vanishes in the middle of its support, meaning~$\widetilde{\rho}_{\lambda_c}(E)$ is singular and $V_{\lambda_c}(x)$ a critical potential. This is an interesting feature that gives rise to universal physics that we shall study more closely in the next subsection. Before doing that, let us work out the deformation of a simple example explicitly.

\subsubsection{Deforming the Gaussian ensemble}

\begin{figure}
\centering
\includegraphics[scale=0.30]{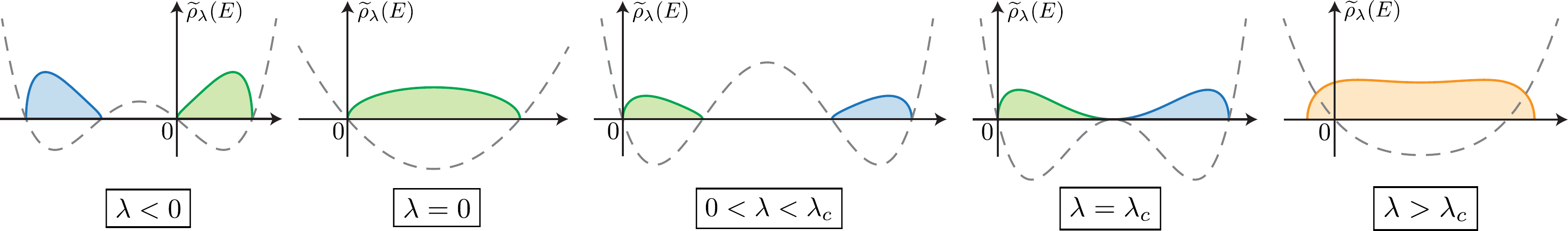}
\caption{Equilibrium spectral density obtained by applying the $T\bar{T}$ deformation on Wigner's semi-circle law for several value of $\lambda$. The green and blue shaded regions correspond to contributions coming from the perturbative and non-perturbative branches respectively. After the transition we observe a single-cut phase in which there is no longer a distinction between the two branches. The dotted line corresponds to the potential $V_\lambda(x)$ in (\ref{eq:24}).}
\label{fig:8}
\end{figure}

Let us consider the simplest Hermitian matrix model, obtained from a Gaussian potential $V(x)$
\begin{equation}\label{eq:27}
V(x)=\frac{8x(x-E_0)}{E_0^2}\ ,
\qquad \qquad
\rho_0(E)=\frac{(4/E_0)^2}{2\pi}
\sqrt{E(E_0-E)}
\times 
\textbf{1}_{[0,E_0]}\ .
\end{equation}
The equilibrium spectral density $\rho_0(E)$ is nothing more than Wigner's semi-circle law centered at~${E_0/2}$. It is straightforward to verify the conditions in (\ref{eq:16}) and (\ref{eq:68}) are satisfied by this potential and spectral density. The $T\bar{T}$ deformation of this model is defined from the potential in~(\ref{eq:81}), that in this case is given by
\begin{equation}\label{eq:24}
V_\lambda(x)=
\frac{8c_\lambda}{E_0^2}
x(1-2\lambda x)
[x(1-2\lambda x)-E_0]\ .
\end{equation}
The spectral density in the double-cut phase is obtained from (\ref{eq:66})
\begin{equation}
\widetilde{\rho}_{\lambda\le \lambda_c}(E)=c_\lambda
\frac{(4/E_0)^2}{2\pi}
|1-4\lambda E|
\sqrt{E(1-2\lambda E)[E_0-E(1-2\lambda E)]}
\times
\big[
\textbf{1}_{[0,E_-(\lambda,E_0)]}+
\textbf{1}_{[E_+(\lambda,E_0),1/2\lambda]}
\big]\ ,
\end{equation}
while in the single-cut phase it is computed from (\ref{eq:16}) as
\begin{equation}\label{eq:26}
\widetilde{\rho}_{\lambda\ge \lambda_c}(E)=
\frac{\left[
2(1-4\lambda E)^2+
8(a_-^2+a_+^2)\lambda^2+8E_0\lambda-3
\right]}
{\pi E_0^2}
\sqrt{(E-a_-)(a_+-E)}
\times
\textbf{1}_{[a_-,a_+]}\ .
\end{equation}
The end points $a_\pm$ are computed from the conditions in (\ref{eq:68}) and given by
\begin{equation}
a_\pm(\lambda,E_0)=
\frac{3\pm \sqrt{6-24\lambda E_0+3
\sqrt{3+(1-16\lambda E_0)^2}}}
{12\lambda}\ .
\end{equation}
In figure \ref{fig:8} we plot the resulting spectral densities and deformed potential (\ref{eq:24}) for several values of $\lambda$, the shaded regions in green and blue corresponding to the contributions from the branches $E_-$ and $E_+$ respectively. After the transition we get the single-cut spectral density (\ref{eq:26}) in orange, where the distinction between the branches is no longer sensible. The support of the spectral density as a function of $\lambda$ is plotted in the right diagram of figure \ref{fig:6}.

\subsection{Critical behavior}
\label{sub:3.2}

We now study the critical behavior of the deformed system as $\lambda\rightarrow \lambda_c$. Let us start by considering the simple Gaussian example (\ref{eq:27}), which is certainly not a critical system since $h(E)$ does not vanish in the support of $\rho_0(E)$. However, after deforming the model and taking $\lambda=\lambda_c$ the spectral density becomes (\ref{eq:29})
\begin{equation}\label{eq:28}
\widetilde{\rho}_{\lambda_c}(E)=
\frac{c_{\lambda_c}}{4\pi}(4/E_0)^2
\left(
\frac{E-2E_0}{2E_0}
\right)^2
\sqrt{
E(4E_0-E)}
\times
\textbf{1}_{[0,4E_0]}\ .
\end{equation}
Since it vanishes at the interior point $2E_0$ where the two cuts merge (see figure~\ref{fig:8}), the potential~${V_{\lambda_c}(x})$~(\ref{eq:24}) is critical. This type of behavior in a Hermitian matrix model was first studied long ago in~\cite{Cicuta:1986pu}, where the transition between single and double-cut phase was shown be be third order. The universal physics associated to the critical behavior is obtained from a standard double scaling limit~\cite{DiFrancesco:1993cyw,Ginsparg:1993is}, first applied to (\ref{eq:28}) in~\cite{Douglas:1990xv}.\footnote{See \cite{Bleher:2002th} for a more recent analysis of the double scaling limit which generalizes to non-symmetric cases.} Physical observables are determined from a solution to a differential equation usually referred as ``string equation", that in this case is given by Painleve II.

The universal characteristics of the system do not depend on the precise details of the model, but only on the rate at which the spectral density vanishes at $2E_0$, quadratic in this case (\ref{eq:28}). The same behavior can be obtained by deforming a different matrix model, as long as the spectral density at~$\lambda_c$ (\ref{eq:29}) satisfies $h(E_0-2\lambda_c E_0^2)=h(E_0)\neq 0$. It is in this sense that the double scaling captures universal features of the transition. What is more, the same critical behavior can be obtained from a different class of matrix model built from unitary instead of Hermitian matrices~\mcite{Periwal:1990gf,Douglas:1990xv,Periwal:1990qb}. In particular, the transition obtained from deforming the Gaussian model (\ref{eq:27}) is in the same universality class as the Gross-Witten third order phase transition in two-dimensional gauge theory~\cite{Gross:1980he}.

This analysis raises the question as to whether we can obtain more general critical behavior, corresponding to $\widetilde{\rho}_{\lambda_c}(E)$ vanishing at a different rate in the interior point $2E_0$. To do so, let us consider a matrix model for which the function $h(E)$ that determines the spectral density (\ref{eq:58}) is given by
\begin{equation}\label{eq:90}
h_k(E)=
b_{k}
\left(
\frac{E_0-E}{E_0}
\right)^{k-1}\ ,
\qquad \qquad
b_k=
\frac{2\pi(k+1)!}{\Gamma(3/2)\Gamma(k+1/2)E_0^2}\ ,
\end{equation}
where $b_k$ is a normalization constant and $k\in \mathbb{N}$. While for $k=1$ we recover the Gaussian example in (\ref{eq:27}), when $k>1$ the system is already critical since $\rho_0(E)$ is supported on $E\in[0,E_0]$ and~${h_k(E_0)=0}$. The double scaling of these family of models was first explored in \cite{Gross:1989aw,Douglas:1989ve,Brezin:1990rb}, where the string equation was shown to be related to the KdV hierarchy. The potential that generates this spectral density is obtained from (\ref{eq:8}) 
\begin{equation}\label{eq:91}
V^{(k)}(x)=
b_k
\sum_{n=0}^{k-1}
\binom{k-1}{n}
\sum_{m=0}^{n+1}
\binom{1/2}{m}\frac{(-E_0)^{m-n}}{(n+2-m)} x^{n+2-m}\ .
\end{equation}
Note that its leading order behavior is given by $V^{(k)}(x)\propto (-x)^{k+1}+\cdots$ with a positive proportionality constant. While this means the system is unstable for $k$ even, it still makes sense as a formal matrix model, see \cite{Eynard:2015aea}. Interestingly, when applying the $T\bar{T}$ deformation we find the associated potential is actually stable for all values of $k$
\begin{equation}
V_\lambda^{(k)}(x)=
c_\lambda V_{\lambda=0}^{(k)}(x-2\lambda x^2)=
\frac{c_\lambda b_k(2\lambda)^{k+1}}
{(k+1)E_0^{k-1}}
x^{2(k+1)}+\cdots\ .
\end{equation}
This gives a nice example in which the deformation cures the inherent instability of the model we started from. The deformed spectral density in the double-cut phase $\lambda\le \lambda_c$ is obtained from $\widetilde{h}_\lambda(E)$ in (\ref{eq:66}), which as we approach $\lambda\rightarrow \lambda_c$ becomes (\ref{eq:29})
\begin{equation}
\widetilde{\rho}_{\lambda_c}^{\,(k)}(E)=
\frac{c_{\lambda_c}}{4\pi}
b_k
\left(
\frac{E-2E_0}{2E_0}
\right)^{2k}
\sqrt{
E(4E_0-E)}
\times
\textbf{1}_{[0,4E_0]}
\ .
\end{equation}
This gives the critical behavior we were after. The spectral density vanishes at the interior point~$2E_0$ at a rate $2k$, generalizing (\ref{eq:28}) beyond quadratic order. The double scaling of this models for arbitrary $k$ was studied in \cite{Crnkovic:1990mr}, where the string equation was shown to be given by the modified KdV hierarchy (mKdV). This is again in the same universality class as critical models built from unitary matrix models \cite{Periwal:1990gf,Periwal:1990qb}. All things considered, applying the $T\bar{T}$ deformation and tuning~${\lambda\rightarrow\lambda_c}$ we get a physical mapping between critical models in the KdV and mKdV hierarchies.

\subsection{Unitary matrices}

Since the $T\bar{T}$ deformation in quantum mechanics is defined in terms of the Hamiltonian operator (which is Hermitian), we have been able to give a natural definition of the deformation for Hermitian random matrices. Generalizing to other matrix ensembles is an interesting question that we adress in this subsection. We use a duality \cite{Dalley:1992br,Mizoguchi:2004ne} between the Hermitian and unitary ensembles to define the $T\bar{T}$ deformation of a unitary matrix model.

The duality between Hermitian and unitary matrices was first noticed in~\cite{Dalley:1992br} for the double scaled models and later generalized in \cite{Mizoguchi:2004ne}. Let us start by showing how the relation works at the level of the matrix partition function. We first write $\mathcal{Z}$ in (\ref{eq:97}) for the Hermitian matrix model in terms of the eigenvalues $\alpha_i\in \mathbb{R}$ of the matrix $M$. After diagonalizing the matrix $M$, standard arguments allow us to write the integral as\footnote{Here and below we are omitting an overall factor of ${\rm Vol}\,(U(N))$ that plays no role in our discussion.}
\begin{equation}
\mathcal{Z}_{\rm hermitian}=
\prod_{i=1}^N
\int_{-\infty}^{+\infty}
d\alpha_i\,
\Delta(\alpha)^2
e^{-\frac{N}{\gamma}V(\alpha_i)}\ ,
\qquad {\rm where} \qquad
\Delta(\alpha)=\prod_{1\le i<j\le N}
(\alpha_j-\alpha_i)
\end{equation}
is the Vandermonde determinant arising from the Jacobian obtained from diagonalization. Changing the integration variable to $\alpha_i=\tan(\theta_i/2)$ with $\theta_i\in(-\pi,\pi]$, $\Delta(\alpha)^2$ transforms in the following way
\begin{equation}
\Delta(\alpha)^2=
\frac{|\Delta(e^{i\theta})|^2}{2^{N(N+1)}}
\prod_{i=1}^N
\cos(\theta_i/2)^{-2(N-1)}
\ .
\end{equation}
Using this, the partition function of the Hermitian matrix model becomes
\begin{equation}\label{eq:43}
\mathcal{Z}_{\rm hermitian}=
\frac{1}{2^{N(N+2)}}
\prod_{i=1}^N
\int_{-\pi}^{\pi}
d\theta_i\,
|\Delta(e^{i\theta})|^2\,
e^{-\frac{N}{\gamma}\left[V(\tan(\theta_i/2))
+\gamma\ln[\cos^2(\theta_i/2)]
\right]}\equiv \frac{\mathcal{Z}_{\rm unitary}}
{2^{N(N+2)}}
\ ,
\end{equation}
where we have defined the partition function of a unitary matrix model built from $U^\dagger=U^{-1}$ as
\begin{equation}\label{eq:45}
\mathcal{Z}_{\rm unitary}=
\int dU
e^{-\frac{N}{\gamma}W(U)}\ ,
\quad {\rm where} \quad
W(U)\equiv V\left[
i\left(\frac{1-U}{1+U}\right)
\right]+\gamma
\ln\left[
\frac{(1+U)^2}{4U}
\right]\ .
\end{equation}
This simple identity allows us to relate expectation value of observables in each theory. For instance, the spectral density $\varrho(\varphi)$ characterizing the eigenvalues $\theta_k$ in the unitary model can be written as
\begin{equation}\label{eq:51}
\begin{aligned}
\varrho(\varphi)\equiv 
\Big\langle
\frac{1}{N}
\sum_{k=1}^{N}\delta(\varphi-\theta_k)
\Big\rangle_{\rm unitary}&=
\mathcal{Z}_{\rm unitary}^{-1}
\int dU\,e^{-\frac{N}{\gamma}W(U)}
\frac{1}{N}
\sum_{k=1}^{N}\delta(\varphi-\theta_k)\\
&=
\mathcal{Z}_{\rm hermitian}^{-1}
\prod_{i=1}^N\int_{-\infty}^{+\infty}
d\alpha_i\,
\Delta(\alpha)^2
e^{-\frac{N}{\gamma}V(\alpha_i)}
\frac{1}{N}
\sum_{k=1}^{N}
\delta(\varphi-2\arctan(\alpha_k))\\
&=
\frac{\big\langle
\frac{1}{N}
\sum_{k=1}^{N}
\delta(\tan(\varphi/2)-\alpha_i)
\big\rangle_{\rm hermitian}}
{2\cos^2(\varphi/2)}
=
\frac{\rho(\tan(\varphi/2))}{2\cos^2(\varphi/2)}\ ,
\end{aligned}
\end{equation}
where $\rho(\cdot)$ is the spectral density of the Hermitian matrix model. In the first line we have used the definition of the expectation value in the unitary matrix model. In the second we used (\ref{eq:43}), written the matrix $U$ integral in terms of its eigenvalues $e^{i\theta_i}$ and changed variables to $\alpha_i=\tan(\theta_i/2)$. In the last line we have used the composition rule of the Dirac delta and reinterpreted in terms of the Hermitian matrix model expectation value. All things considered, we get a simple relation between the spectral densities of each model. Other observables can be related in a similar fashion.

\begin{figure}
\centering
\includegraphics[scale=0.27]{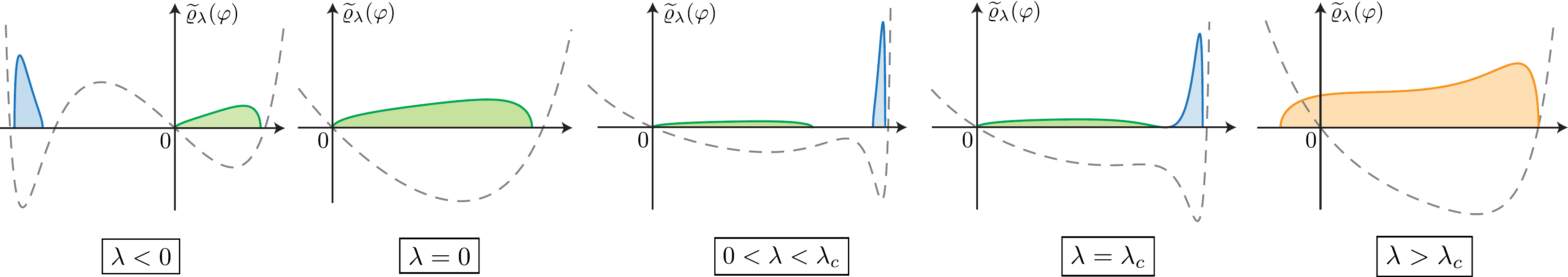}
\caption{Equilibrium spectral density and potential obtained by $T\bar{T}$ deforming the unitary matrix model dual to the Gaussian Hermitian matrix model in (\ref{eq:27}). The spectral density $\varrho(\varphi)$ is defined in the compact domain $\varphi\in[-\pi,\pi]$. The overall behavior of the model is analogous to the Hermitian matrix model in figure \ref{fig:8}.}
\label{fig:2}
\end{figure}

Let us now use this duality to define the $T\bar{T}$ deformation for a unitary matrix model. Using the formula for the deformation of the Hermitian model potential (\ref{eq:81}) and the relation in (\ref{eq:45}), we find
\begin{equation}\label{eq:50}
W_\lambda(U)=
c_\lambda
V\left[i\left(\frac{1-U}{1+U}\right)+2 \lambda
\left(\frac{1-U}{1+U}\right)^2\right]+
\gamma\ln\left[
\frac{(1+U)^2}{4U}
\right]\ .
\end{equation}
To write the deformation without needing to refer to the potential $V(M)$ in the Hermitian model, we can rewrite the right-hand side using (\ref{eq:45})
\begin{equation}\label{eq:52}
W_\lambda(U)=
c_{\lambda}
W(f(U))+
\gamma\left\lbrace
\ln\left[\frac{(1+U)^2}{4U}\right]
-c_\lambda\ln\left[
\frac{(1+f(U))^2}{4f(U)}
\right]
\right\rbrace
\  ,
\end{equation}
where we have defined
\begin{equation}
f(y)\equiv \frac{(1+y)^2}{(1+y)-i\lambda(1-y)^2}-1\ .
\end{equation}
This defines the deformation of the unitary matrix model. Note that when~${\lambda=0}$ we have~${c_{\lambda=0}=1}$,~${f(U)=U}$ and the additional terms in~(\ref{eq:52}) drops out.

As a simple example, we can take the Gaussian ensemble in the Hermitian matrix model~(\ref{eq:27}) and study its $T\bar{T}$ deformation from the perspective of the unitary matrix model. Using~(\ref{eq:50}) and~(\ref{eq:51}) we compute and plot the equilibrium density and potential in figure \ref{fig:2} for several values of~$\lambda$. The overall behavior is analogous to that of the Hermitian matrix model, shown in figure \ref{fig:8}, with the difference that $\varphi\in[-\pi,\pi]$.

\section{Discussion}
\label{sec:5}

In this final section we study the $T\bar{T}$ deformation of double scaled matrix models, as well as perform some quantitative comparisons between matrix models and finite cut-off JT gravity results.

\subsection{Double scaled models}
\label{subsec:5.1}

To illustrate how a double scaled model can be obtained from matrix model, let us start by considering the simple Gaussian example in (\ref{eq:27}). The double scaled model is completely characterized by its spectral density $\rho_{\rm ds}(E)$, which can be obtained from $\rho_0(E)$ in (\ref{eq:27}) from the following limit
\begin{equation}\label{eq:92}
\rho_{\rm ds}(E)\equiv
\lim_{E_0\rightarrow +\infty}E_0^{3/2}\rho_0(E)=
(8/\pi)
\sqrt{E}\times
\textbf{1}_{[0,+\infty)}\ ,
\end{equation}
where the power $E_0^{3/2}$ is chosen so as to pick up the leading order behavior in the $E_0$ expansion. This limit corresponds to zooming to the edge of the spectral density $\rho_0(E)$ at $E=0$.\footnote{The limit in (\ref{eq:92}) is a simple way of obtaining the spectral density of the double scaled model. However, we should keep in mind that the appropriate way of taking the double scaling limit of a matrix model involves a precise procedure, see \cite{DiFrancesco:1993cyw,Ginsparg:1993is} for reviews on the methods.} Since~$\rho_{\rm ds}(E)$ is not normalizable, the essential feature in (\ref{eq:92}) is its square root behavior with respect to the energy. This simple double scaled model, sometimes called the ``Airy model", belongs to a larger family characterized by~${\rho_{\rm ds}(E)\propto E^{k-1/2}}$ for $k\ge 1$, obtained from double scaling the critical matrix models studied in subsection \ref{sub:3.2}.

While for any matrix model there is a single associated double scaled model, the inverse is not true. There are an infinite number of matrix models that give rise to the same double scaled model. This is particularly clear from (\ref{eq:92}), as one can easily construct other matrix model spectral densities~$\rho_0(E)$ which have the single square root leading order behavior in the large $E_0$ limit. In short, there is no universal way of reversing the double scaling limit. This implies that in order to define the $T\bar{T}$ deformation of a double scaled model, using the definition for the matrix model studied in section \ref{sec:3} and given by (\ref{eq:31}) is not particularly useful, as the result would be by construction non-universal. Therefore, we must start again from the quantum mechanical definition in section~\ref{sec:2}. In particular we can use the expression for the deformed spectral density in (\ref{eq:21}) or (\ref{eq:22}), apply them to $\rho_{\rm ds}(E)$ and study the deformation from there.

This is the approach we take here. In doing so, there are several issues and ambiguities we must deal with. While we are not able to provide definite answers to all of the problems, we discuss and provide partial answers that we hope are valuable for future investigations on this subject. We shall drop the notation $\rho_{\rm ds}(E)$, understanding that if the support of a spectral density is unbounded, it corresponds to a double scaled model.

\subsubsection{Negative deformation coupling}

Let us start by considering the deformation of double scaled models with a negative deformation coupling $\lambda<0$. In this regime we do not have to deal with the complexification of energy eigenvalues, which occurs at $\lambda>\lambda_c=1/8E_0=0$ since $E_0\rightarrow \infty$. From the quantum mechanical analysis we can define the deformed double scaled model by considering either a single (\ref{eq:21}) or both branches (\ref{eq:22}) that solve the flow equation 
\begin{equation}\label{eq:89}
\begin{aligned}
{\rm Single\,\,branch}:&
\qquad
\rho_\lambda(E)=
(1-4\lambda E)\rho_0(E-2\lambda E^2)\times
\textbf{1}_{[0,+\infty)}\ ,
\\[5pt]
{\rm Both\,\,branches}:&
\qquad
\widetilde{\rho}_\lambda(E)=c_\lambda
|1-4\lambda E|\rho_0(E-2\lambda E^2)\times
\big[
\textbf{1}_{[0,+\infty)}+
\textbf{1}_{(-\infty,1/2\lambda]}
\big]\ .
\end{aligned}
\end{equation}
While the spectral density $\rho_\lambda(E)$ has the usual support on the positive real line, $\widetilde{\rho}_\lambda(E)$ is instead non-zero in two disjoint semi-infinite segments, which makes it harder to interpret as a double scaled model.\footnote{It might be possible to study $\widetilde{\rho}_\lambda(E)$ from the more abstract topological expansion of \cite{Eynard:2007kz}.} Since $\rho_\lambda(E)$ in (\ref{eq:89}) has the standard support on the positive real line, we shall explore the $T\bar{T}$ deformation defined from including the single branch, using the description of the model as a particular combination of multi-critical  models. We shall not review the basics of this formalism but point the interested reader to the reviews in \cite{Ginsparg:1993is,DiFrancesco:1993cyw} and the more recent applications in \cite{Johnson:2019eik,Johnson:2020heh,Johnson:2020exp,Johnson:2020mwi,Okuyama:2019xbv,Okuyama:2020ncd,Okuyama:2020qpm,Johnson:2020lns}.

The double scaled models can be studied in a perturbative expansion for a small parameter $\hbar$ that play the same role as the $1/N$ in the ordinary matrix model expansion. The spectral density~$\rho(E)$ to all orders in $\hbar$ can be computed as \cite{Gross:1989aw,Banks:1989df,Ginsparg:1993is}
\begin{equation}\label{eq:100}
\rho(E)=\int_{-\infty}^0dx\,|\psi_E(x)|^2\ ,
\qquad \qquad
\mathcal{H}[u]=-\hbar^2\partial_x^2+u(x)\ ,
\end{equation}
where $\mathcal{H}\psi_E(x)=E\psi_E(x)$. The central object in this formalism is the potential $u(x)$ that is determined from a differential equation called the ``string equation", which for a single-cut Hermitian matrix model is given by
\begin{equation}\label{eq:96}
\mathcal{R}\equiv\sum_{k=1}^{\infty}
t_k\widetilde{R}_k[u]+x=0\ .
\end{equation}
Here $\widetilde{R}_k[u]$ is a $k$-th order polynomial in $u(x)$ and its derivatives defined by Gel'fand-Dikii but normalized so that the coefficient $u^k$ is unity.\footnote{See \cite{Gelfand:1975rn} for more details and explicit expressions.} The double scaled model is essentially defined by the coefficients $t_k$ appearing in $\mathcal{R}$ (\ref{eq:96}). Once these are fixed, we can solve the differential equation~${\mathcal{R}=0}$ for $u(x)$, compute the spectrum of the operator $\mathcal{H}[u]$ in (\ref{eq:100}) and obtain the full spectral density~$\rho(E)$. This means that in order to define the $T\bar{T}$ deformation of the model we must find the following map
\begin{equation}
T\bar{T}\,\,{\rm deformation}:
\qquad
t_k
\qquad \longrightarrow \qquad
t_k(\lambda)\ .
\end{equation}

To leading order in $\hbar$, the string equation $\mathcal{R}_0=\lim_{\hbar\rightarrow 0}\mathcal{R}$ is related to $\rho_0(E)$ through the following relation derived in \cite{Johnson:2020lns}
\begin{equation}\label{eq:111}
\mathcal{R}_0=
\sum_{k=1}^{\infty}t_ku_0^k+x=2\hbar
\int_0^{u_0}\frac{dE}{\sqrt{u_0-E}}\rho_0(E)+x\ .
\end{equation}
Using that the deformed spectral density is given in the first line of (\ref{eq:89}) we get
\begin{equation}\label{eq:94}
\mathcal{R}_0(\lambda)\equiv 2\hbar
\int_0^{u_0}
\frac{dE}{\sqrt{u_0-E}}
(1-4\lambda E)
\rho_0(E-2\lambda E^2)+x\ .
\end{equation}
For any particular model we can solve the integral, expand  in a power series in $u_0$ and identify the deformed coefficients $t_k(\lambda)$. To do so, let us assume the undeformed spectral density can be expanded in the following way
\begin{equation}\label{eq:36}
\rho_0(E)=
\frac{1}{2\pi \hbar}
\sum_{q=1}^{\infty}
a_q
E^{q-1/2}\ ,
\end{equation}
for some coefficients $a_q\ge 0$. Inserting this in (\ref{eq:94}) we can exchange the integral with the series, since all the terms that are being integrated are non-negative. Each of the integrals can be solved to give a hypergeometric function, so that we get the following expansion for $\mathcal{R}_0(\lambda)$
\begin{equation}\label{eq:37}
\mathcal{R}_0(\lambda)=
\sum_{q=1}^{\infty}
\frac{2(2q-1)!}
{4^{q}q!(q-1)!}
a_q
u_0^q
\,_2F_1\left[
\frac{1+2q}{-2},
\frac{3+2q}{2},
1+q,2\lambda u_0
\right]
+x\ .
\end{equation}
Since $\lambda<0$ and $u_0\ge 0$ the hypergeometric functions in each term are real. This is not the case if we where to naively take $\lambda$ positive in this expression.

To identify the deformed coefficients $t_k(\lambda)$ we must write $\mathcal{R}_0(\lambda)$ as a power series expansion in~$u_0$, as done in (\ref{eq:111}). When doing so we encounter an issue, since the series expansion of the hypergeometric function has a finite radius of convergence, given by $u_0<1/|2\lambda|$. This is a problem, as the coefficients $t_k(\lambda)$ obtained in this way are not going to describe the physics for arbitrarily high energies, but up to a maximum energy $E_{\rm max}(\lambda)$ given by
\begin{equation}\label{eq:3}
E_{\rm max}(\lambda)=\frac{1}{2|\lambda|}\ .
\end{equation} 
While for higher energies the leading string equation in (\ref{eq:37}) is still well defined through analytic continuation, the coefficients $t_k(\lambda)$ are not. It is interesting that although for $\lambda<0$ there is no issue with the complexification of the spectrum (\ref{eq:71}), a very different phenomenon forces us to introduce a truncation in the energy.

Keeping this in mind, we can compute $t_k(\lambda)$ using the standard expansion of the hypergeometric function around the origin, together with the Cauchy product for infinite series:
\begin{equation}\label{eq:38}
\boxed{t_k(\lambda)=
t_k(0)
(k+1/2)
\sum_{q=0}^{k-1}
\frac{(-2\lambda)^q}{q!}
\frac{a_{k-q}}{a_k}
\frac{\Gamma(k+1/2-q)}{\Gamma(k+3/2-2q)}}
\ ,
\end{equation}
where the undeformed coefficients are
\begin{equation}
t_k(0)=
\frac{2a_k(2k-1)!}{4^kk!(k-1)!}\ .
\end{equation}
This transformation defines the $T\bar{T}$ deformation of the double scaled model.

\paragraph{Airy model:} Let us start by considering the simple Airy model, in which the leading spectral density is given by $\rho_0(E)=\sqrt{E}/2\pi \hbar$. The deformed coefficients $t_k(\lambda)$ in (\ref{eq:38}) are easily computed and given by
\begin{equation}
t_k(\lambda)=
\frac{(-2\lambda)^{k-1}}{2k!(k-1)!}
\frac{\Gamma(k+3/2)}{\Gamma(7/2-k)}\ .
\end{equation}
While for $\lambda=0$ the only non-vanishing coefficient is $t_1$, when we turn on the deformation we have an infinite number of higher order contributions. The leading string equation is given by
\begin{equation}
\mathcal{R}_0(\lambda)=\sum_{k=1}^{\infty}t_k(\lambda)u_0^k+x=
\frac{1}
{2}
u_0
\,_2F_1\left[-
\frac{3}{2},
\frac{5}{2},
2,2\lambda u_0
\right]
+x=0\ .
\end{equation}

\paragraph{JT gravity:} A more interesting example is obtained from the double scaled model that describes JT gravity. The leading spectral density in this case is given by \cite{Saad:2019lba}
\begin{equation}
\rho_0(E)=\frac{\sinh(2\pi\sqrt{E})}
{4\pi^2 \hbar}=
\frac{1}{2\pi \hbar}
\sum_{q=1}^{\infty}
\frac{(2\pi)^{2(q-1)}}
{(2q-1)!}E^{q-1/2}\ .
\end{equation}
Identifying the coefficients $a_q$ we can compute $t_k(\lambda)$ in (\ref{eq:38}) and find it can be written in terms of a hypergeometric function
\begin{equation}
t_k(\lambda)=
\frac{\pi^{2(k-1)}}{2 k!(k-1)!}
\,_3F_0\left[
1-k,\frac{2k+1}{-4},\frac{2k-1}{-4};;\frac{8\lambda}{\pi^2}
\right]\ .
\end{equation}
When $\lambda=0$ the hypergeometric function goes to one and we identify the prefactor as the undeformed coefficients of JT gravity \cite{Okuyama:2019xbv,Johnson:2019eik}. For non-zero $\lambda$ and fixed $k$ the hypergeometric function is a simple polynomial in $\lambda$ of order $(k-1)$.

\subsubsection{Positive deformation coupling}

Let us now consider the $T\bar{T}$ deformation for positive coupling $\lambda$, which turns out being quite different. Consider the simplest family of double scaled models obtained from multi-critical potentials labeled by $k\in \mathbb{N}$
\begin{equation}\label{eq:95}
\rho_0(E)=E^{k-1/2}\times \textbf{1}_{[0,+\infty)}\ .
\end{equation}
Deforming this spectral density according to (\ref{eq:21}) or (\ref{eq:22}) is not as straightforward as for $\lambda$ negative, as in this case we have to deal with the complexification of the spectrum. Moreover, double scaled models do not seem to have the structure that allowed us in section \ref{sec:3} to deal with this issue in a natural and unique way through a phase transition. Due to the lack of a better procedure, we shall introduce a truncation in the spectrum of (\ref{eq:95}), given by $E_{\rm max}(\lambda)=1/4\lambda$. Doing so, the deformed spectral density including either a single (\ref{eq:21}) or both branches (\ref{eq:22}) is given by
\begin{equation}\label{eq:115}
\begin{aligned}
{\rm Single\,\,branch}:&
\qquad
\rho_\lambda(E)=
\left(\frac{2k+1}{a^{2k+1}}\right)
(a-E)
[E(2a-E)]^{k-1}
\sqrt{E(2a-E)}
\times
\textbf{1}_{[0,a]}\ ,
\\[5pt]
{\rm Both\,\,branches}:&
\qquad
\widetilde{\rho}_\lambda(E)=
\left(\frac{2k+1}{2a^{2k+1}}\right)
|a-E|
[E(2a-E)]^{k-1}
\sqrt{E(2a-E)}
\times
\textbf{1}_{[0,2a]}\ ,
\end{aligned}
\end{equation}
where we have defined $a=1/4\lambda$. Both of these expressions have been rescaled in order to yield a normalized spectral density supported on a finite interval.

The expressions in (\ref{eq:115}) are quite interesting, as they seem to correspond to the large $N$ spectral densities of a matrix model \textit{without} double scaling. As we have previously explained, inverting the double scaling is a non-universal procedure, meaning there are an infinite number of ways of doing so. Still, the $T\bar{T}$ deformation is selecting a particular way of inverting the double scaling limit in~(\ref{eq:95}). This resonates with the effect of the deformation on two-dimensional QFTs by an irrelevant operator, as in that case the deformation also picks a particular trajectory in the renormalization group flow, among many possibilities.

With this in mind, let us inspect more closely the two expressions in (\ref{eq:115}), whose main difference is their support. We first compare with the spectral density of a general large $N$ matrix model with polynomial potential, given in (\ref{eq:61}). Note that only $\widetilde{\rho}_\lambda(E)$ in (\ref{eq:115}) has the appropriate structure, and we can easily identify the polynomials $\widetilde{h}_\lambda(E)$ and $\widetilde{\sigma}_\lambda(E)$ that characterize the model. Due to the indicator function, this is not the case for $\rho_\lambda(E)$ in (\ref{eq:115}). The next natural step is to use the general conditions in (\ref{eq:68}) to compute the potential $V_\lambda(x)$ that generates the spectral density $\widetilde{\rho}_\lambda(E)$, similarly as done previously in section \ref{sec:3}. However, we stumble into a problem, since the second relation in (\ref{eq:68}) can never be satisfied, i.e.
\begin{equation}\label{eq:11}
\underset{z=+\infty}{\rm Res}
\left[
\widetilde{h}_\lambda(z)
\sqrt{\widetilde{\sigma}_\lambda(z)}
\right]=
2\pi
\left(\frac{2k+1}{2a^{2k+1}}\right)
\underset{z=+\infty}{\rm Res}
\left[
(a-z)[z(2a-z)]^{k-1}
\sqrt{z(z-2a)}
\right]
=0\neq -2\ .
\end{equation}
Recall that this constraint comes from requiring the simple condition that the resolvent $R(z)$ in (\ref{eq:12}) behaves like $R(z)=1/z+\cdots$ for large $z$. The vanishing of the residue in (\ref{eq:11}) implies the leading behavior $1/z$ vanishes. Overall, this means that even though $\widetilde{\rho}_\lambda(E)$ has the appropriate structure, it does not arise from the large $N$ limit of a matrix model with a \textit{polynomial} potential $V_\lambda(x)$. 

What about non-polynomial potentials? It is still possible the spectral densities in (\ref{eq:115}) correspond to a matrix model with a more complicated potential. To determine this, we can use the following expression that relates the potential $V(x)$ of a matrix model with its equilibrium density
\begin{equation}\label{eq:101}
V'(x)=2
\fpint{{\rm supp}\,\rho}{\,}
\frac{\rho_0(E)}{x-E}dE\ ,
\qquad {\rm for}\,\,x\in {\rm supp}\,\rho\ ,
\end{equation}
where the integral is computed in the principal value regularization. This follows from the saddle point analysis of the partition function \cite{Anninos:2020ccj} and only assumes the potential $V(x)$ is a well behaved function so that the matrix integral converges (see section 3.2 in \cite{Anninos:2020ccj}). As an example, it is straightforward to consider~$\rho_0(E)$ in the Gaussian example (\ref{eq:27}), solve the integral and find
\begin{equation}\label{eq:102}
V_{\rm gaussian}'(x)=
\frac{16}{\pi E_0^2}
\fpint{0}{E_0}
\frac{\sqrt{E(E_0-E)}}{x-E}dE
=\frac{8(2x-E_0)}{E_0^2}\ ,
\end{equation}
which gives the known answer given in (\ref{eq:27}). Note that even though the formula (\ref{eq:101}) only determines the potential in the region ${x\in {\rm supp}\,\rho}$, in this case the expression is naturally extended to the whole real line $x\in \mathbb{R}$.

Let us now consider the $T\bar{T}$ deformed spectral densities in (\ref{eq:115}). For any value of $k$ the integral can also be solved explicitly, although the final answer is much more complicated. As an example, let us consider $\rho_\lambda(E)$ in (\ref{eq:115}) with $k=1$. The potential obtained from (\ref{eq:101}) is given by
\begin{equation}\label{eq:103}
V^{(k=1)}_\lambda(x)=
-\pi\bar{x}^3
+(2+3\pi)\bar{x}^2
-
\frac{1}{2}(8+3\pi)\bar{x}
-1
+4[\bar{x}(2-\bar{x})]^{3/2}
{\rm arccoth}\left[
\sqrt{2/\bar{x}-1}
\right]+2\ln(1-\bar{x})\ ,
\end{equation}
where $\bar{x}=x/a$. This expression is clearly quite complicated and not polynomial. Moreover, while for $x\in[0,a]$ the potential is real (as required), it cannot be extended to the whole real line $x\in\mathbb{R}$, as done for the simple Gaussian case (\ref{eq:102}), since (\ref{eq:103}) becomes complex for $x>a$. Overall, it is unclear whether the deformed spectral densities in (\ref{eq:115}) can be made sense of as a random matrix model.

\subsection{Finite cut-off JT gravity}
\label{subsec:5.2}

As mentioned in the introduction, recent work has shown interesting connections between the $T\bar{T}$ deformation with positive coupling~$\lambda$ and finite cut-off holography \cite{McGough:2016lol}. For two-dimensional JT gravity this was explored in \cite{Iliesiu:2020zld}, where the finite cut-off disc partition function was computed and matched with the $T\bar{T}$ deformation of the Schwarzian quantum mechanics \cite{Gross:2019ach}. Since higher topology contributions in ordinary JT gravity are captured by a double scaled Hermitian matrix model \cite{Saad:2019lba}, is there a deformed matrix model that captures higher topology contributions of finite cut-off JT gravity?


To answer this, let us consider the matrix expectation value of double trace operators in the large~$N$ limit, which take a universal and particularly simple form. The connected expectation value of two resolvent insertions 
\begin{equation}\label{eq:105}
R_0(z_1,z_2)\equiv
\lim_{N\rightarrow \infty}
\frac{1}{N^2}
\Big\langle
{\rm Tr}
\frac{1}{z_1-M}
{\rm Tr}
\frac{1}{z_2-M}
\Big\rangle_c\ ,
\end{equation}
for a single-cut matrix model only depends on the endpoints of the interval $(a_-,a_+)$ where the spectral density is supported. It is given by~\cite{Eynard:2015aea}
\begin{equation}\label{eq:108}
R_0(z_1,z_2)=
\frac{-1}{2(z_1-z_2)^2}\left[
1+
\frac{(a_-+a_+)(z_1+z_2)/2-(a_-a_++z_1z_2)}
{\sqrt{(z_1-a_-)(z_1-a_+)}
\sqrt{(z_2-a_-)(z_2-a_+)}}
\right]\ ,
\end{equation}
which in the coincident limit $z_1=z_2=z$ becomes
\begin{equation}\label{eq:107}
R_0(z,z)=
\frac{(a_+-a_-)^2}
{16(z-a_+)^2(z-a_-)^2}\ .
\end{equation}
This provides a simple expression that we can use to compare with finite cut-off JT gravity results.

To do so, we use the dictionary that allows us to translate gravitational to matrix model observables \cite{Saad:2019lba}. The gravitational partition function in JT gravity~$Z(\beta)$ with a single asymptotic boundary of renormalized length $\beta$ is identified with the following operator insertion in the matrix model
\begin{equation}\label{eq:113}
\langle {\rm Tr}\,e^{-\beta M} \rangle
\qquad \longleftrightarrow \qquad
Z(\beta)
\ .
\end{equation}
Adding more boundaries to the gravitational path integral corresponds to additional insertions of~${{\rm Tr}\,e^{-\beta M}}$. Using this, we have the following identification with $R_0(z_1,z_2)$ in (\ref{eq:105}) 
\begin{equation}\label{eq:104}
R_0(z_1,z_2)
\qquad \longleftrightarrow \qquad
\int_{0}^{+\infty}
d\beta_1d\beta_2
Z_{\rm cylinder}(\beta_1,\beta_2)e^{\beta_1z_1+\beta_2z_2}
\ ,
\end{equation}
where $Z_{\rm cylinder}(\beta_1,\beta_2)$ is the leading genus contribution to the gravitational path integral with two asymptotic boundaries, i.e. cylinder topology. The integral transform in $\beta_i$ is required in order to change the insertion of the exponential matrices $e^{-\beta M}$ in (\ref{eq:113}) to resolvents (\ref{eq:105}).

To compute the right hand side, we use some finite cut-off results obtained in \cite{Iliesiu:2020zld}. Using the decomposition of multi-boundary surfaces developed in \cite{Saad:2019lba}, the cylinder partition function is constructed from the trumpet partition function $Z_{\rm trumpet}$, that contains a geodesic boundary of length~$b$ and a boundary of \textit{finite} length $L$\footnote{This is obtained from equation (4.8) in \cite{Iliesiu:2020zld} after including the boundary counterterm $e^{-L\phi_b}$ and rescaling by an overall factor so that we recover the trumpet partition function of ordinary JT gravity \cite{Saad:2019lba} in the appropriate limit. Below we comment on some subtle aspects regarding the derivation of this result.\label{foot}}
\begin{equation}\label{eq:72}
\begin{aligned}
Z_{\rm trumpet}=
e^{-L\phi_b}
L\phi_b
\frac{J_1\big(\phi_b\sqrt{b^2-L^2}\big)}{\sqrt{b^2-L^2}}=
e^{-L\phi_b}
(L\phi_b/b)
\sum_{n=0}^{\infty}
\left[
\frac{L^2\phi_b}
{2b}
\right]^n
\frac{J_{n+1}(b\phi_b)}{n!}
\ ,
\end{aligned}
\end{equation}
where $\phi_b$ is the value of the dilaton at the boundary and in the second equality we have used the identity in equation (8.515) of \cite{gradshteyn2007}. Instead of working with the parameters $(\phi_b,L)$ it is convenient to use~$(a,\beta)$ with $a=1/4\lambda$, where $\lambda$ would be the $T\bar{T}$ deformation parameter of the matrix model. We can translate between these quantities using the identifications given in~\cite{Iliesiu:2020zld}~${(\phi_b,L)=\sqrt{a/2}\,(1,2\beta)}$, which ensures the matching between the disc partition function and the $T\bar{T}$ deformation of the Schwarzian quantum mechanics. 

To compute the cylinder partition function we must take two different values of $\beta_1$ and $\beta_2$ while $a_1=a_2=a$, since from the boundary perspective the matrix model is deformed by the single parameter $\lambda$. The cylinder partition function is then obtained by gluing two trumpets and integrating over all possible values of $b\in \mathbb{R}_+$ using the Weil-Petersson measure $db\,b$ \cite{Saad:2019lba}
\begin{equation}\label{eq:73}
\begin{aligned}
Z_{\rm cylinder}(a,\beta_1,\beta_2)&=
\int_0^{+\infty}db\,b\,
Z_{\rm trumpet}(b,a,\beta_1)
Z_{\rm trumpet}(b,a,\beta_2)\\
&=
\frac{1}{2}
e^{-(\beta_1+\beta_2)a}
\sum_{n,m=0}^{\infty}
\frac{(\beta_1 a)^{2n+1}}
{n!2^{n}}
\frac{(\beta_2 a)^{2m+1}}
{m!2^{m}}
\int_0^{+\infty}
dy
\frac{J_{n+1}(\sqrt{y})}
{y^{\frac{n+1}{2}}}
\frac{J_{m+1}(\sqrt{y})}
{y^{\frac{m+1}{2}}}\\
&=
e^{-a(\beta_1+\beta_2)}
\sum_{n,m=0}^{\infty}
\frac{2\beta_1^{2n+1}\beta_2^{2m+1}}
{(n!m!)^2(1+n+m)}
\left(\frac{a}{2}\right)^{2(n+m+1)}\ ,
\end{aligned}
\end{equation}
where we have changed integration variables to $b=\sqrt{2y/a}$ and used the series representation in (\ref{eq:72}) to solve the integral. We should be careful with this expression, as we have carelessly exchanged the integral and infinite series. To check that no issue arises from this technicality, we can take $\beta_1=\beta_2$ where the series can be solved and written in terms of modified Bessel functions
\begin{equation}
Z_{\rm cylinder}(a,\beta,\beta)=
\frac{1}{2}e^{-2a\beta}(a\beta)^2\left[
I_0(a\beta)^2-I_1(a\beta)^2
\right]\ge 0\ .
\end{equation}
This agrees with result obtained from directly solving the $b$ integral in (\ref{eq:73}) after using $Z_{\rm trumpet}$ as written in the first expression in (\ref{eq:72}). For $\beta_1\neq \beta_2$ we have directly solved the integral numerically, compared with the truncated series (\ref{eq:73}) and found agreement to arbitrary precision. Overall, this means we can trust the series expansion in (\ref{eq:73}) for the cylinder partition function.

We can now insert $Z_{\rm cylinder}$ in (\ref{eq:104}) and compute the integral for each of the terms in the series. Exchanging the series with the integral is fully justified in this case, as each term in the series is non-negative. In this way, we can write (\ref{eq:104}) as
\begin{equation}
R_0(z_1,z_2)
\qquad \longleftrightarrow \qquad
\sum_{n,m=0}^{\infty}
\frac{(2n+1)!(2m+1)!}
{(n!m!)^2(1+m+n)}
\frac{2(a/2)^{2(n+m+1)}}
{(a-z_1)^{2(n+1)}(a-z_2)^{2(m+1)}}
\ ,
\end{equation}
where $R_0(z_1,z_2)$ is given in (\ref{eq:108}). The comparison of these quantities is simpler when $z_1=z_2=z$, were the series can be solved and we find
\begin{equation}
R_0(z,z)=
\frac{(a_+-a_-)^2}
{16(z-a_+)^2(z-a_-)^2}
\qquad \longleftrightarrow \qquad
\frac{a^2}{4z^2(2a-z)^2}\left[
2-\frac{a^2}{(a-z)^2}
\right]
\ .
\end{equation}
As a check, both side match perfectly for the ordinary JT gravity after taking $a=1/4\lambda\rightarrow \infty$ and~${(a_-,a_+)=(0,\infty)}$. However, for finite cut-off JT gravity (corresponding to $a$ finite), there are no values of $a_\pm$ we can take so that both expressions agree. This shows the computation of the finite cut-off observables using the decomposition of the surfaces developed in \cite{Saad:2019lba} does not yield a result compatible with a random matrix model. A different approach must instead be developed for computing higher genus finite cut-off observables in JT gravity. 

Some readers might think this conclusion is too abrupt. For instance, one can consider the possibility that finite cut-off JT gravity is described by a multi-cut instead of a single-cut matrix model. However, this does not seem to be possible, since it is well known large observables of multi-cut matrix models do not have a an well defined large $N$ limit \cite{Bonnet:2000dz}. For instance, while $R_0(z_1,z_2)$ can still be computed explicitly for double-cut matrix models, the answer depends on $N$ non-analytically, i.e. it depends on whether~$N$ is even or odd (see equation (3.18) in~\cite{Bonnet:2000dz}).
 
One can also look more closely at the computation of the trumpet partition function (\ref{eq:72}), obtained from solving the Wheeler-de Witt equation \cite{Iliesiu:2020zld}. In a similar way as there are two branches $E_\pm(\lambda,E)$ solving the $T\bar{T}$ flow equation (\ref{eq:30}), there are two independent solutions to the Wheeler-de Witt equation. The trumpet partition function in (\ref{eq:72}) is obtained by taking a particular combination between these two solutions, corresponding to the following two terms in the integral
\begin{equation}\label{eq:114}
Z_{\rm trumpet}=
\int_0^{\phi_b^2}
dE
\frac{\cos(b\sqrt{E})}{2\pi\sqrt{E}}
\left[
e^{-L\left[\phi_b-\sqrt{\phi_b^2-E}\right]}
-
e^{-L\left[\phi_b+\sqrt{\phi_b^2-E}\right]}
\right]\ ,
\end{equation}
which is equivalent to (\ref{eq:72}). While in \cite{Iliesiu:2020zld} this particular combination is well motivated, it seems reasonable to explore other combinations, which essentially means replacing the minus in the second term (\ref{eq:114}) by an arbitrary parameter $q$. However, when doing so and using the result to compute the cylinder partition function as the first line in~(\ref{eq:73}), one finds $Z_{\rm cylinder}$ is finite only when $q=-1$. This supports the expression for the trumpet partition function (\ref{eq:72}), as computed in \cite{Iliesiu:2020zld}.

\bigskip
\leftline{\bf Acknowledgments}
\noindent I am thankful to Joaquin Turiaci for comments on the draft, and Bertrand Eynard and Taro Kimura for correspondence. I am particularly grateful to Clifford Johnson for guidance and collaboration on the initial stages of this project. This work is  supported by the DOE grant DE-SC0011687 (USC), NSF grant PHY-1748958 (KITP) and the Heising-Simons Foundation.

\bibliography{sample}
\bibliographystyle{JHEP}

\end{document}